\documentclass{elsart}
\usepackage[dvips]{epsfig}
\usepackage{latexsym,fancybox,pifont}
\begin{document}
\begin{frontmatter}
\title{Bootstrap quark model and spectra masses, electromagnetic
  properties of low-lying hadrons}
\author{S.M.~Gerasyuta, and S.~Bielefeld}
\address{Dept. of Physics, Technical Academy, St.Petersburg,\\ 
 Max-Planck-Institut f\"ur Kernphysik, Heidelberg}
\date{\today}
\begin{abstract}
The paper is devoted to the construction of low-energy
quark scattering amplitudes. Bootstrap quark model allows to describe
the light and heavy hadron spectroscopy based on three principles:
unitarity, analicity and crossing symmetry. The resulting quark
interaction appeared to be effectively short-range. This interaction
is determined mainly by the exchange in the gluon channel: the
constituent-gluon mass appeared to not be small. Our calculation
indicates an important role of an interaction which is induced by
instantons. Such an interaction is necessary for deriving both the
pion mass and $\eta$-$\eta'$ mass splitting. However, due to the rules
of $1/N_c$-expansion this interaction influences slightly the other
channels while in the $\eta$-$\eta'$ channel it provides the correct
values of masses and gives the angle of $\eta_1-\eta_8$ mixture close
to that of the quark model. We discuss the possibility of constructing
the amplitudes which take into account the quark confinement.
\end{abstract}
%
\tableofcontents
\end{frontmatter}
\section{Introduction}
  The quark phenomenology is based on the hypothesis of the
  quasinuclear hadron structure: the hadron consists of two (meson)
  or three (baryon) spatially separated dressed quarks like the
  nucleus consists of nucleons. A dressed quark of the quasinuclear
  model should be a quark-gluon cluster consisting of the valence
  quark, sea quarks and anti-quarks. The small size of the dressed
  quark may be connected with the small range of gluon
  interaction. One should try to describe soft processes operating
  with  dressed quarks as quasi-particles. The project is devoted to
  the construction of low-energy quark scattering amplitudes. The
  iteration bootstrap procedure includes two types of point-like input
  interactions, namely the four-fermion interaction with quantum
  numbers of the gluon and four-fermion interaction induced by
  instantons (the exchange of white states with
  $J^P=0^{\pm}$). Calculations performed in the framework of this model
  argue in favor of the quasinuclear quark structure of hadrons. The
  resulting quark interaction appeared to be effectively
  short-range. This interaction is determined mainly by the exchange
  in the gluon channel: the calculated constituent gluon mass
  $M_G=0.67~{\rm GeV}$ to be not small. The creation of mesons (pion
  included) is mainly due to the gluon exchange. The mass values of
  the lowest mesons ($J^P=0^{-+},1^{--},0^{++}$) and their quark
  content are obtained. In the color channel $\bar{3}_c$ the bound
  states of the scalar diquarks with the masses $m_{ud}=0.72~{\rm GeV}$ and
  $m_{us}=m_{ds}=0.86 GeV$ are obtained. The dispersion relation
  technique allows us to calculate the form factors of light mesons
  using quark amplitudes in corresponding channels. The calculated
  values of the charge radii squared of pion and Kaon are in good
  agreement with data. The calculated values of the charge radii
  squared of u, d, s quarks are equal: ~$r_u^2=2.06~{\rm GeV}^{-2},
  r_d^2=1.75~{\rm GeV}^{-2}, ~r_s^2=1.03~{\rm GeV}^{-2}$. The charge radii of
  non-strange and strange scalar diquarks are calculated:
  $\langle r_{ud}^2\rangle^{1/2}=0.55~{\rm fm}$,~$\langle
  r_{us}^2\rangle^{1/2}=0.65~{\rm fm}$,~$\langle
  r_{ds}^2\rangle^{1/2}=0.5~{\rm fm}$.\\  
  In this consideration
  one has neglected the forces of confinement. This is due to the fact
  that the model is actually based on the assumption that the
  confinement radius is much larger than the radius of dressed quarks,
  and also larger than the radius of the forces responsible for the
  existence of the low-lying hadrons. In the case of the orbital
  excited mesons: P-,~D-,~F-wave mesons one can not neglect the
  confinement potential. Therefore the main idea is the modification
  of precise theory by help of changing the interaction of quarks and
  anti-quarks at large range and do not charge the interaction at small
  range. Therefore one constructs the approximate model, in which
  there are free quarks, but the theory of quark-anti-quark scattering
  uses only the necessary part of the confinement potential. This idea
  can be used in different approach. In our case the confinement
  potential is imitated by the simple increasing of constituent quark
  masses. It allows us to construct the P-,~D-,~F-wave meson
  amplitudes and calculate the mass spectrum, which is in good
  agreement with data. We calculate the behavior of meson Regge
  trajectories for the low-energy region, which is based on the
  principles of multi-color QCD.
  In the framework of an approach developed previously for light
  quarks one calculates the scattering amplitudes of heavy quarks
  $q\bar{Q}-q\bar{Q}$ and $Q\bar{Q}-Q\bar{Q}$ (q=u,~d,~s; Q=c,~b). The
  interaction of heavy quarks is described by forces corresponding of
  exchange of light white and colored mesons. The main role in the
  formation of the spectrum of heavy mesons is played by the forces
  that correspond to exchange of a massive gluon. It is necessary to
  take into account the renormalization of the quark amplitudes for
  light quarks, which is determined by the rescattering of the heavy
  quarks. We obtained the masses of lowest multiplets of c- and
  b-mesons with quantum numbers $J^{PC}=0^{-+},1^{--},0^{++}$. \\
  Then we
  use a quark interaction potential, which is obtained from the
  nonrelativistic limit of relativistic quark amplitudes of the
  Bootstrap quark model. But the quark amplitudes obtained in the
  Bootstrap procedure depend not only on the square t of the momentum
  transfer, but also on the energy variable
  $s$. Therefore, a literal
  transition to nonrelativistic potentials is not possible: these
  amplitudes rather correspond to quasipotentials. If the energy
  $s=s_0$ is fixed, and
  already at fixed energy the dependence on the 
  momentum transfer is considered to be a potential nature.\\ 
  We investigated S-wave baryons with quantum numbers
  $J^P=\frac{1}{2}^{+},\frac{3}{2}^{+}$ consisting of u and d
  quarks. The masses and wave function of the baryons were obtained
  from the solution of the nonrelativistic Faddeev equations for the
  three-quark system with the diquark potentials $V_{0^+}$ and
  $V_{1^+}$. 
  One obtained the electric and magnetic form factors of
  the nucleons. The values of the masses of the nucleons and the
  $\Delta$ isobar agree well with the experimental values, then we
  have a good agreement with experiment for the magnetic moments
  $\mu_p=2.79 (2.79)$ and $\mu_n=-1.86 (-1.91)$. The charge radius of
  the proton has been to be almost a factor of two smaller than
  experimental value $\langle R_p^2\rangle^{1/2}=0.4 (0.8)~{\rm fm}$, charge
  radius of the neutron has been found to be zero. This calculation has
  not new parameters besides the Bootstrap quark model results. The
  behavior of additional linear potential are investigated and do not
  essentially change the results. In our paper a relativistically
  generalization of the three-body Faddeev equations was obtained in
  the form of dispersion relations in the pair energy of two
  interacting particles. The mass spectrum of S-wave baryons
  $J^P=\frac{1}{2}^+,\frac{3}{2}^+$ including u,~d,~s quarks was
  calculated by a method based on isolating of the leading
  singularities in the amplitude. We searched for the approximate
  solution of integral three-quark equations by taking into account
  two-particle and triangle singularities, all the weaker ones being
  neglected. If we considered the approximation, which corresponds to
  taking into account two-body and triangle singularities, and define
  all the smooth functions of the sub-energy variables (as compared
  with the singular part of the amplitude) in the middle point of
  physical region of Dalitz-plot, then the problem reduces to one of
  solving simple algebraic system equations. The masses of the baryons
  of the two lowest multiplets with $J^P=\frac{1}{2}^+$ and
  $\frac{3}{2}^+$ are calculated and found to be in a good agreement
  with the experimental results. The behavior of the electromagnetic
  form factors of nucleons and hyperons in the region of low and
  intermediate momentum transfers $Q^2 < 1.5~{\rm GeV}^2$ is determined. The
  calculated value (without new parameters) of the charge radius of
  the proton was found to be $\langle R_p^2\rangle^{1/2}=0.4~{\rm fm}$,
  hyperons $\langle R_{\Sigma^+}^2\rangle^{1/2}=0.43~{\rm fm}$, $\langle
  R_{\Sigma^-}^2\rangle^{1/2}=0.39~{\rm fm}$, $\langle
  R_{\Xi^-}^2\rangle^{1/2}=0.38~{\rm fm}$. The charge radius of the neutron,
  and other noncharge hyperons were found to be practically equal to
  zero. In the framework of dispersion relation technique the
  relativistic Faddeev equations for charmed baryons are found. The
  approximate solutions of the relativistic three-particle problem
  based on the extraction of leading singularities of amplitudes are
  obtained. The mass values of lowest charmed baryon multiplets
  $J^P=\frac{1}{2}^+,\frac{3}{2}^+$ are calculated. The behavior of
  electromagnetic form factors of charmed baryons
  ($J^P=\frac{1}{2}^+$) in the region of low and intermediate momentum
  transfers $Q^2 < 1.5~{\rm GeV}^2$ is determined. The calculated value
  (without new parameters) of the charge radii baryons like
  $\Sigma_{c}^{++}$, $\Lambda_{c}^+$ and so on $(0.12 - 0.20)~{\rm fm}$. The
  charge radii of noncharge charmed baryons are equal to zero. We
  calculated the multiplets masses of radial excited baryon, included
  the Roper resonance, with quantum number
  $J^P=\frac{1}{2}^+,\frac{3}{2}^+$ at $N=2,56^{\star}$. The problem
  of confinement is solved analogous to the case of excited
  mesons. Finally, a practical treatment of relativistic two- and
  three-hadron quark systems have been developed. The theory is based
  on the three principles of unitarity, analiticity and crossing
  symmetry. These principles are applied to the two-body subenergy
  channels. The scattering quark amplitudes allow to calculate the
  spectrum and electromagnetic properties of hadrons. 
In soft processes, where small momentum transfers are essential, the
perturbative QCD technique is not applicable. In this case the way of
phenomenology should include quark model results as well. 
Quark models make it possible to describe the qualitative properties
of QCD in the nonperturbative region. In the framework of quark models
important information is obtained on the properties of light
mesons~\cite{1}-~\cite{9} and baryons~\cite{10}-~\cite{16}. The reason
for the successful use of quark potential models is connected with a
successful choice of the effective parameters: the mass of the dressed
quarks, the characteristics of the confinement potential, and the
coupling constant $\alpha_s$. In quark models, which describe rather
well the masses and static properties of light hadrons, the masses of
the quarks usually have the same values for both mesons and
baryons. However, this is achieved at the expense of some difference
in the characteristic of the confinement potential. It should be borne
in mind that for a fixed hadron mass the masses of dressed quarks
which enter into the composition of the hadron will become smaller
when the slope of the confinement potential increases or its radius
decreases.\\ 
Therefore, conversely one can change the masses of the dressed quarks
when going from the spectrum of s-wave mesons to s-wave baryons, while
keeping the characteristic of the confinement potential unchanged.
The mass spectrum of charmed and beauty mesons is subject of
investigation of many theoretical papers, based on various
approaches. Calculation in potential models are
widespread~\cite{17}-~\cite{22}.\\
The unquestionable advantage of the potential approach to the
description of heavy mesons is its simplicity and transparency. It
makes it possible to calculate (in the framework of the adopted model
of the potential) the positions of the levels, the widths of radiative
transitions between levels, and the widths of annihilation
decays. Thus, in fact it succeeds in determining of behavior of the wave
function of the system in a fairly wide range of distances, which is
very important for an understanding of the dynamics of the interaction
of the quarks. A serious difficulty of the potential approach is the
fact that it cannot be fully substantiated in the framework of
QCD. The method of QCD sum rules has been successfully and to
determine the masses of the light and heavy mesons~\cite{23,24}. In
this approximation the vacuum condensates are considered as
phenomenological parameters determined from the experimental data or
from self-consistency of the sum rules. The most serious difficulty
for the application of the sum-rule method to bottomonium (for
instance) is the problem of taking into account relativistic
corrections. So far there has been no such calculation.

\section{Iteration method in the Bootstrap procedure}
The project is devoted to the construction of low-energy quark scattering
amplitudes. The amplitudes of dressed quarks are calculated in the
framework of the dispersion technique with the help of the iteration
Bootstrap procedure. The theory is based on three principles of
unitarity, analyticity and crossing symmetry which are applied to the
three-body sub-energy channels.\\
The iteration Bootstrap procedure for quark-quark and quark-anti-quark
amplitudes was made with the four-fermion interaction as an input:
\begin{equation}
\label{eq:1}
g_\nu(\bar{q}\gamma_\mu\vec{\lambda}q)(\bar{q}\gamma_\mu\vec{\lambda}q)
\end{equation}
where $\vec{\lambda}$ are the Gell-Mann matrices. The point-like
structure of this interaction is motivated by the idea of the two
characteristic sizes in the hadron. On the other hand, the
applicability of~(\ref{eq:1}) is verified by the success of the
De-Rujula-Georgi-Glashow quark model~\cite{1}, where only the
short-range part of the Breit potential connected with the gluon
exchange is responsible for the mass splitting in hadron
multiplets. However, the correct description of low-lying meson masses
is impossible using the effective gluon exchange
interaction~(\ref{eq:1}) only. The large value of the $\eta - \eta'$ mass
splitting and the small pion values indicate the existence of an
additional interaction. Such an additional four-fermion interaction
can be induced by instantons. This interaction should also be a
short-range one because of the small radius of the effective gluon
interaction. We have included this type of interaction in the
iteration Bootstrap calculation. The scheme of the iteration Bootstrap
procedure suggested in this paper is as follows. The partial
amplitudes are calculated through the dispersion technique.\\
The zero approximation in Fig. 1d is drawn as the sum of
diagrams shown in Fig. 1a,~b,~c etc. The amplitude of the first
approximation (Fig. 1h) is obtained when the zero approximation
amplitude is taken as an interaction like Eq.(\ref{eq:1}), i.e. the
interaction in the first approximation is determined by the exchange
diagrams of the zero approximation. The first approximation amplitude
is the sum of diagrams shown in Figs. 1e,~f,~g etc. The use of the
first approximation amplitude as an ``interaction force'' provides us
with the second approximation, and so on. We consider the rescattering
amplitudes in three sub-energy channels u,~s,~t.
\begin{figure}
\begin{center}
\unitlength1cm   
\begin{minipage}[t]{90mm}
\epsfxsize=90mm\epsfbox{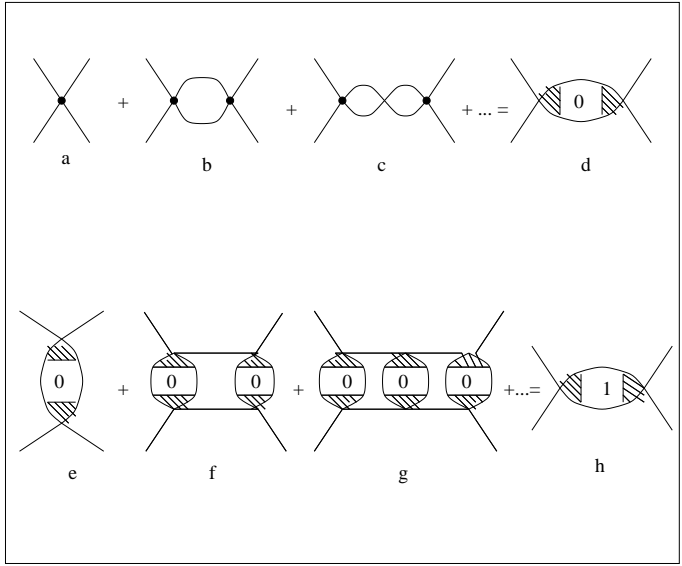}
\end{minipage}
\caption{\label{fig:1}Iteration of Bootstrap procedure}
\end{center}
\end{figure}
We construct scattering amplitudes of the dressed quarks of three
flavors (u,~d,~s) the poles of these amplitudes determine the masses
of the light mesons~\cite{25,26}. The masses of the constituent quarks
u and d are of the order of 300-400 MeV; strange quark is 100-150 MeV
heavier. The constituent quark is a color triplet and quark amplitudes
obey the global color symmetry.
\begin{table}[!hhh]
\begin{center}
\begin{tabular}{cccccc}
    & $J^{PC}=0^{-+}$&   &$J^{PC}=1^{--}$& & $J^{PC}=0^{++}$ \\\hline
$\pi$ & 0.14~(0.14)  &$\rho$ & 0.77~(0.77)& $a_0$ & 0.78~(0.98) \\
$\eta$ & 0.48~(0.55)  &$\omega$ & 0.77~(0.78)& $f_0$ & 0.87~(0.98) \\ 
$K$ & 0.50~(0.50)  &$K^*$ & 0.89~(0.89)& $K_0^*$ & 0.88~(1.35) \\
$\eta'$ & 0.96~(0.96)  &$\phi$ & 1.00~(1.02)& $f_0$ & 1.16~(1.30) \\
& $\theta=-24^{\circ}$& & $\theta=30^{\circ}$&
&$\theta=-82^{\circ}$
\end{tabular}
\caption{\label{tab:1} Masses of the lowest meson multiplets (GeV) - 
Experimental values of mesons are in parenthesis. Parameters of the
model: gluon constant $g_V=0.226$, cut-off $\Lambda_q=17.3$, instanton
constants $g_I=-0.081$, $g_S=0.55$, $m_u=0.385$, $m_s=0.501$~GeV.}
\end{center}
\end{table}
For the $0^+$ multiplet the discrepancy between calculated and
observed values of masses more than for $0^{-+}$ and $1^{--}$. It is
possible that this is due to the admixture of the glueball states as
$q\bar{q}q\bar{q}$-states in the scalar mesons. In all the versions of
our calculus there is a bound state in the gluon channel with the mass
of order of 0.7 GeV. Its quark wave function is \[(1.06 (u\bar{u}+d\bar{d}) +0.87 ~s\bar{s})/\sqrt{3};\] hence with a good
accuracy it is a singlet of the flavor $SU(3)_f$--group. This bound
state should be identified as a constituent gluon. In the diquark
channel there are bound states with $c=\bar{3}$ and $J^P=0^+$.\\
The dispersion relation technique allows us to calculate the form
factors of composite particles using the quark amplitudes in
corresponding channels. The calculated values of both pion and Kaon
charge radii are equal
\begin{equation} 
\langle R_{\pi^{\pm}}^2\rangle^{1/2}=0.76 ~{\rm fm}\hspace{2em}
\langle R_{K^{\pm}}^2\rangle^{1/2}=0.69 ~{\rm fm}\hspace{2em}
R_{K_0}^2=-1.1 ~{\rm GeV}^{-2}
\end{equation}

For the interaction with the photon current dressed-quark radii are
determined by the behavior of quark amplitudes in the channel $1^-$
near $t=0$. The calculated values of the charge radii are equal for
u,~d,~s quarks:
\begin{equation}
\langle r_{ud}^2\rangle^{1/2}= 0.55 ~{\rm fm} \hspace{2em}
\langle r_{us}^2\rangle^{1/2}= 0.65 ~{\rm fm} \hspace{2em}
\langle r_{ds}^2\rangle^{1/2}= 0.5 ~{\rm fm}   
\end{equation}

Parameters of this model are: cut-off energy $\Lambda=17.3$, gluon
constant $g_V=0.226$, four-fermion instanton interaction $g_I=-0.081$,
interaction with exchange of white isosinglet $g_S=0.55$, quark masses
$m=0.385 ~{\rm GeV}$, $m_s=0.501 ~{\rm GeV}$, see Table~\ref{tab:1}.

\section{P-,~D-,~F-wave mesons in the Bootstrap method}
One of the most surprising, and so far unexplained, dynamical features
of the spectroscopy of light quarks is the existence of five almost
completely filled nonets in the quark-anti-quark $q\bar{q}$-system,
predicted in the nonrelativistic quark model: two S-wave and three
P-wave nonets. 
In the well established $J^{PC}=1^{--}$ and $2^{++}$ nonets ideal
mixing of the $SU(3)_f$ singlet and isoscalar components of the octet
is realized, and this implies the suppression of the transition
$(u\bar{u}+d\bar{d})/\sqrt{2}\leftrightarrow s\bar{s}$ in these
channels (the Okubo-Zweig-Iitzuka-rule). We should say that the
simplest nonrelativistic quark model predicts ideal nonets. The status
of the axial with $J^{PC}=1^{++}$ and $1^{+-}$ is not as clear as the
status of the tensor mesons. Apparently, the $1^{++}$ nonet is also an
ideal nonet. The $1^{+-}$ nonet is not yet completely filled, with
eight of the ten members established. 
The strange meson $K_1^*$, belonging to the multiplet with
$J^{PC}=1^{++}$, is observed in a mixture with the meson $K_1$ of the
multiplet with $J^{PC}=1^{+-}$. Experimentally, the $K_1^*$ meson is
observed with the main decay channel Kp, and $K_1$ is observed with
the main channel $K^* \pi$.\\
The pseudoscalar $0^{-+}$ nonet is not ideal. Nowadays, the reason for
this is understood. It is connected with the solution of the
U(1)-problem in QCD, and with instanton
contributions~\cite{28}-~\cite{31}. At present the properties of the
scalar mesons with$J^{PC}=0^{++}$ are among the most obscure aspects
of the spectroscopy of light mesons. Although the resonances
$a_0$,~$f_0$,~$K_0^*$ and $\tilde{f}_0$ can be considered well
established, their interpretation in the language of the quark model
is not clear. The superficially natural assumption that they form a
nonet of $q\bar{q}$ states cannot be reconciled with their
masses. This may be due to the existence of an appreciable admixture
of glueball states or multiquark states in the scalar
mesons~\cite{32,33}.
In this section the Bootstrap quark model technique allows us to
calculate the spectra mass of P-,~D-,~F-mesons. The Regge trajectories
for the low-energy region, which describe the mesonic resonances with
orbital numbers $L=0,1,2,3$ are constructed. \\
The technique of carrying out the Bootstrap procedure of excited
mesonic states is analogous to one of low-lying mesons. However, the
forces of the interaction between quarks are effectively of short
range. The effective short-range nature is a consequence of two
circumstances. Exchange of a gluon state is most important in the
interaction of the quarks. A constituent gluon is a fairly massive
particle (700 MeV), and this is one of the reasons for the short-range
nature. The other circumstance is that the quark amplitude depend
relatively weak on the energy in the region $t < 0$. Since the values
of the amplitudes of the crossed channels for $t< 0$ are precisely the
interaction forces, relatively small variation of quark amplitude
in the region $t<0$ lead to an extra decrease of the range of
interaction of the quarks. This is connected with the assumption that
the confinement radius is much larger than the radius of the forces
responsible for the existence of the low-lying hadrons, i.e. the
confinement forces can be neglected. But when one  researches the
excited states the confinement potential can not be neglected. The
main idea is the modification of precise theory by help of changing
the interaction of quarks and anti-quarks at large range and do not
change the interaction at small range. Therefore one constructs the
approximate model in which there are free quarks, but the low-lying
spectrum of mesons is former. \\
In the framework of this model we can construct the theory of
quark-anti-quark scattering using only the 
necessary part of the confinement potential. In our case the
confinement potential is imitated by the simple increasing of
constituent quark masses. The shift of quark mass (parameter
$\Lambda$) effectively takes into account the changing of the
confinement potential. We have shown that inclusion of only gluon
exchange indeed does not lead to the appearance of bound states
corresponding to the mesons of the P-,~D-,~F-waves. The using of mass
shift $\Lambda_I$ is possible to obtain the mass spectra of
P-,~D-,~F-mesons (Tables \ref{tab:2},~\ref{tab:3},~\ref{tab:4})~\cite{34,35}.  
In Table \ref{tab:5} we use the results of Bootstrap quark model with
u,~d,~s--quarks. 
\begin{table}
\begin{center}
\begin{tabular}{cccccc}
    & $J^{PC}=1^{++}$&   &$J^{PC}=1^{+-}$& & $J^{PC}=2^{++}$ \\\hline
$a_1$ & 1.273~(1.260)  &$b_1$ & 1.203~(1.235)& $a_2$ & 1.320~(1.320) \\
$f_1$ & 1.273~(1.285)  &$h_1$ & 1.203~(1.170)& $f_2$ & 1.320~(1.270) \\ 
$K_1^*$ & 1.385~(1.400)&$K_1$ & 1.308~(1.270)& $K_2$ & 1.436~(1.430) \\
$f_2$ & 1.497~(1.420)  &$h'_1$ & 1.414~(-)& $f_2$ & 1.552~(1.525) 
\end{tabular}
\caption{\label{tab:2}Masses of three P-wave multiplets (GeV) - 
  $\Delta_P=0.275$ GeV, other parameter as for Table~\protect{\ref{tab:1}}.}
\end{center}
\end{table}

\begin{table}
\begin{center}
\begin{tabular}{cccc|cccc}
    & $J^{PC}=1^{--}$&    &$J^{PC}=2^{-+}$ & & $J^{PC}=2^{--}$&
    &$J^{PC}=3^{--}$ \\\hline 
$\rho_1$ & 1.590~(1.700)  &$\pi_2$ & 1.620~(1.670) & $\rho_2$ &
    1.670~(-)     &$\rho_3$ & 1.690~(1.690) \\ 
$\omega_1$ & 1.590~(1.600)&$\eta_2$ & 1.620~(-) & $\omega_2$ &
    1.670~(-)   &$\omega_3$ & 1.690~(1.670)\\  
$K_1^*$ & 1.690~(1.680)   &$K_2$ & 1.720~(1.710) & $K_2^*$ &
    1.780~(1.820)  &$K_3^*$ & 1.800~(1.780) \\ 
$\phi_1$ & 1.800~(1.680)  &$\eta_2$ & 1.840~(-) & $\phi_2$ & 1.900~(-)
    &$\phi_3$ & 1.920~(1.850) 
\end{tabular}
\caption{\label{tab:3}Masses of D-wave meson multiplets (GeV)
  - $\Delta_D=0.460$ GeV, other parameter as for
  Table~\protect{\ref{tab:1}}.} 
\end{center}
\end{table}

\begin{table}
\begin{center}
\begin{tabular}{cccc|cccc}
    & $J^{PC}=2^{++}$&&$J^{PC}=3^{+-}$
    &&$J^{PC}=3^{++}$&&$J^{PC}=4^{++}$ \\\hline 
 $a_2$ & 1.930~(-)  &$b_3$ & 1.950~(-) &$a_3$ & 1.960~(-)
&$a_4$ & 2.050~(-) \\ 
 $f_2$ & 1.930~(-)  &$h_3$ & 1.950~(-) &$f_3$ & 1.960~(-)
&$f_4$ & 2.050~(2.050)\\   
 $K_2^*$ & 2.030~(-)&$K_3$ & 2.060~(-) &$K_3^*$ & 2.070~(-)
&$K_4^*$ & 2.160~(2.045) \\ 
 $f_2$ & 2.140~(-)  &$h_3$ & 2.170~(-) &$f_3$ & 2.180~(-)
&$f_4$ & 2.280~(-) 
\end{tabular}
\caption{\label{tab:4}Masses of F-wave meson multiplets (GeV)
  -- $\Delta_F=0.640$ GeV, other parameter as for
  Table~\protect{\ref{tab:1}}.}
\end{center}
\end{table}

\begin{table}
\begin{center}
\begin{tabular}{ccc}
Regge trajectories $\alpha(t)$ & $\alpha(0)$ & $\alpha'$~${\rm GeV}^{-2}$ \\\hline
$\alpha_{\rho,\omega}$ & 0.5 (0.5) & 0.9 (0.9) \\
$\alpha_{K^*}$         & 0.4 (0.4) & 0.8 (0.8) \\
$\alpha_{\varphi}$     & 0.2 (0.1) & 0.8 (0.9) \\
$\alpha_{\pi}$         & 0.  (0.)  & 0.8 (0.8) \\
$\alpha_{K}$           & -0.3 (-0.3) & 0.7 (0.7) \\
$\alpha_{\eta}$        & -0.2 (-0.2) & 0.8 (0.8) \\
$\alpha_{\eta'}$       & -0.6 (-)  & 0.8 (-) \\
$\alpha_{a_0,f_0}$     & -0.3 (-0.5) & 0.6 (0.6) \\
$\alpha_{K_0^*}$       & -0.4 (-)  & 0.6 (-) \\
$\alpha_{\tilde{f}_0}$ & -0.6 (-)  & 0.6 (-) \\
$\alpha_{a_1,f_1}$     & -0.4 (-)  & 0.8 (-) \\
$\alpha_{K_1}$         & -0.5 (-0.5) & 0.7 (0.7) \\
$\alpha_{\tilde{f}_1}$ & -0.6 (-)  & 0.7 (-) 
\end{tabular}
\caption{\label{tab:5}Regge trajectories $q\bar{q}$ mesons --
  $\alpha(t)=\alpha_0+\alpha' t$; here, we use the results of
  Bootstrap quark model with u,~d,~s-quarks. Experimental results are
  in parenthesis.}
\end{center}  
\end{table}

In the recent papers~\cite{36,37} the high energy asymptotic of
multicolor QCD is considered, that allows to construct the
Bethe-Salpeter equation for the n reggeized gluons. Then the author
obtained the Pomeron trajectory as the bound state of two reggeized
gluons. Therefore it is important to obtain the meson Regge
trajectories of glueballs. 

\section{Heavy mesons ($J^{PC}=0^{+-},1^{--},0^{++}$)}
We consider the spectrum of the lowest $Q\bar{Q}$ and $Q\bar{q}$
mesons constructed from light (q=u,~d,~s) and heavy (Q=c,~b,~t)
quarks. The main role in the formation of the spectrum of heavy mesons
is played by the forces that correspond to exchange of a massive
gluon. Owing to the point-like nature of the interaction of heavy
quarks, their contribution leads only to numerical renormalization of
the loop-diagrams of light quarks. Therefore, the introduced vertex
function of the interaction of heavy quarks can effectively take into
account his renormalization. The parameters $\alpha_Q$ and
$\alpha_{Qq}$ of the interaction of heavy quarks and cut-off
parameters $\Lambda_Q$ and $\Lambda_{Qq}$ are  given by a definite set
of experimental values of $Q\bar{Q}$ and $Q\bar{q}$-mesons.
\begin{figure}[!h]
\begin{center}
\unitlength1cm   
\begin{minipage}[t]{90mm}
\epsfxsize=90mm\epsfbox{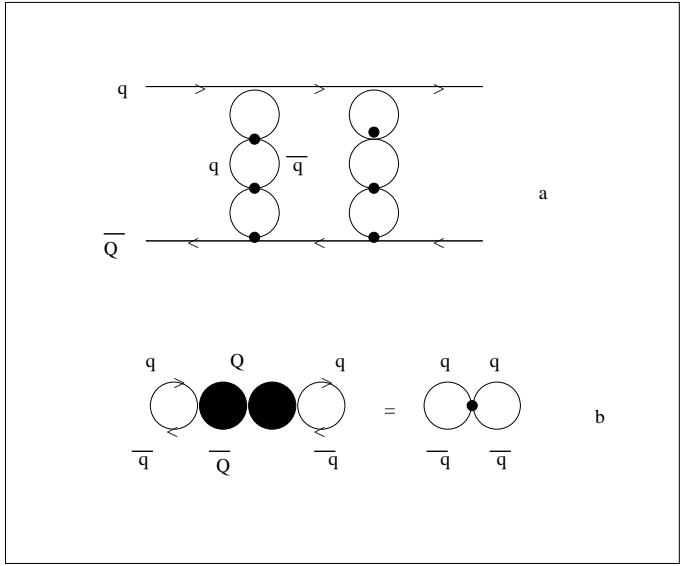}
\end{minipage}
\end{center}
\caption{\label{fig:2}(a) Diagrams which determine the interaction
  amplitudes of heavy quarks, (b) renormalization of the vertex of the
  interaction of light quarks due to the contribution of heavy quarks.}
\end{figure}

\begin{figure}
\begin{center}
\unitlength1cm   
\begin{minipage}[t]{90mm}
\epsfxsize=90mm\epsfbox{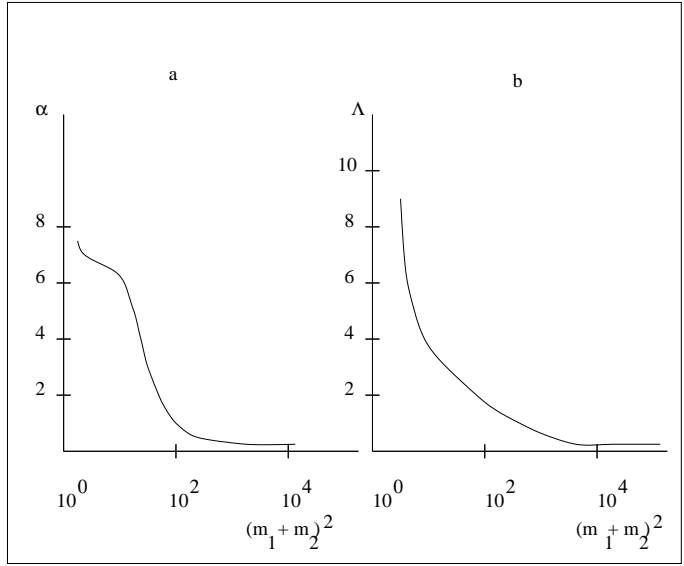}
\end{minipage}
\end{center}
\caption{\label{fig:3}Approximation curve of the variation vertex
  function (a), and the cut-off (b) with increasing mass~\protect{\cite{38}}.}
\end{figure}

We have obtained the masses of mesons with c and b quarks\\
($J^{PC}=0^{-+}$,$0^{++}$,$1^{--}$). In the calculation we have used
the parameters 
$\alpha_Q$ and $\alpha_{Qq}$, and also the parameter $\Lambda_Q$, the
cut-off in the scattering amplitude of heavy quarks. The cut-off in
the scattering amplitude of heavy quarks with light quarks has been
chosen in the following form:
$\Lambda_{Qq}=\frac{1}{4}(\sqrt{\Lambda_Q}+\sqrt{\Lambda_q})^2$. The
value $\Lambda_q=17.3$ and masses $m_{u,d}=385 ~{\rm MeV}$, $m_s=501
~{\rm MeV}$.
The parameters $\alpha_Q$, $\alpha_{Qq}$ and $\Lambda_q$ have been
determined for the c quark from the experimental values of the masses
of the meson $D$, $D^*$, and $J/\Psi$, and for the b quark from the
masses of the mesons $B$, $B^*$, and $\Upsilon$. The masses of the
heavy quarks are also parameters of the model and have been determined
from the best description of the masses of the $\eta_c$ for the c
quark and $B_s$ for the b quark. The results of the calculation of the
meson mass spectrum are  given in Tables \ref{tab:6}
and~\ref{tab:7}~\cite{38}. Fig.~\ref{fig:3} shows the approximation
curve of the variation of the vertex function in (a) and that of the
cutoff in (b) with increasing mass~\cite{38}. 
\begin{table}[!h]
\begin{center}
\begin{tabular}{cccccc}
     & M($0^{-+}$)   &        & M($1^{--}$) &  & M($0^{++}$) \\\hline
$D$  &   1.867 (1.867)&$D^*$   &2.010 (2.010)&$c\bar{u}$&2.119 (-) \\
$D_s$&   2.010 (1.971)&$D^*_s$ &2.120 (2.113)&$c\bar{s}$&2.300 (-) \\
$\eta_c$&2.955 (2.980)&$J/\Psi$&3.097 (3.097)&$\chi_0$  &3.453 (3.415)
\end{tabular}
\caption{\label{tab:6}Masses of the lowest states of charmonium and of
  states with open charm (GeV) - Note: the parameters of the model
  are: $\Lambda_c=5.94$, $\alpha_{qc}=6.56$, $\alpha_c=5.41$ and
  $m_c=1.645 ~GeV$.- In parenthesis we give the experimental numbers.}
\end{center}
\end{table}
\begin{table}[!hh]
\begin{center}
\begin{tabular}{cccccc}
     & M($0^{-+}$)   &        & M($1^{--}$) &  & M($0^{++}$) \\\hline
$B$  &  5.270 (5.270)&$B^*$   &5.320 (5.320)&$b\bar{u}$&5.486 (-) \\
$B_s$&  5.375 (5.340)&$B^*_s$ &5.425 (5.390)&$b\bar{s}$&5.652 (-) \\
$b\bar{c}$&6.085 (-) &$b\bar{c}$&6.320 (-)&$b\bar{c}$&6.735 (-)\\
$b\bar{b}$&9.340 (-) &$\Upsilon$&9.460 (-)&$b\bar{b}$&10.070 (-)
\end{tabular}
\caption{\label{tab:7}Masses of the lowest states of bottomonium and of
  states with open bottom (GeV) - Note: the parameters of the model
  are: $\Lambda_b=4.91$, $\alpha_{qb}=4.08$, $\alpha_{cb}=2.44$,
  $\alpha_b=1.12$ and $m_b=4.940 ~{\rm GeV}$.- In parenthesis we give the
  experimental numbers.} 
\end{center}
\end{table}

       Our model is actually based on the assumption that the confinement
radius is much larger than the radius of dressed quarks, and also
larger than the radius of forces responsible for the existence of the
low-lying hadrons. In the discussion of heavy mesons we have neglected
the forces on confinement. However, estimates show that for meson of
type $\bar{q}t$ the confinement radius becomes comparable with the
radius of gluon exchange.

\section{Bootstrap quark potential and nonrelativistic Faddeev
  equations. Electromagnetic properties of non-strange baryons}
\begin{figure}
\begin{center}
\unitlength1cm   
\begin{minipage}[t]{90mm}
\epsfxsize=90mm\epsfbox{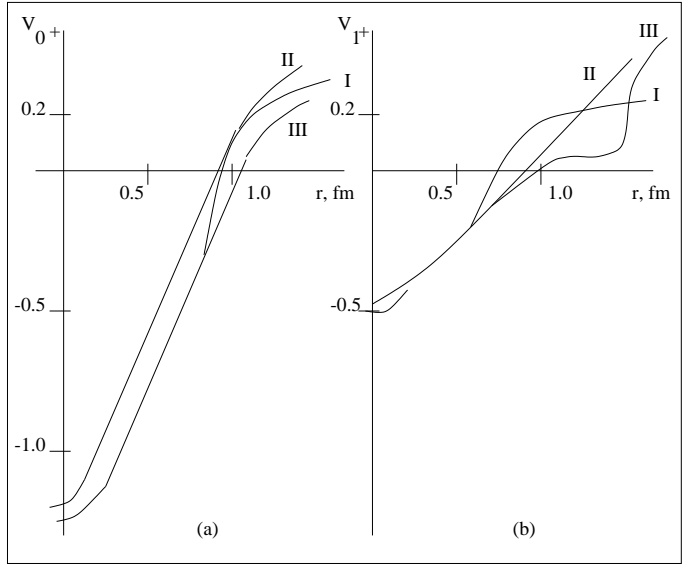}
\end{minipage}
\end{center}
\caption{\label{fig:4}(a) Quark interaction potential in the state
  with $J^P=0^+$; three versions (I)-(III) of the choice of the
  confinement potential $r_0=0.6,1.0,1.3$ fm, respectively. - (b)
  Quark interaction potential in state $J^P=1^+$.}
\end{figure}

With a sufficient number of parameters the potential models give a
good description of the spectrum of hadrons. However, a serious
difficulty for potential scattering is that its characteristics cannot
be calculated directly in the framework of QCD. If the gluon field is
unable to follow a change of the quark system, there is retardation,
and, as a consequence, the interaction acquires a non-potential nature.
Under these conditions a successful description of hadron in
the framework of the potential approach seems highly nontrivial. It is
probable that the above-mentioned non-potential nature shows up only at
intermediate distances. At short distances the Coulomb-like one-gluon
exchange potential works, and at large distances a linear potential is
applicable. In a purely potential approach the non-potential nature of
the interaction at intermediate distances effectively reduces to a
redefinition of the parameters of the potential, which in this
approach are not calculated, but fitted for optimal agreement with the
experimental data. \\
We use a quark interaction potential, which is obtained from the
nonrelativistic limit of relativistic quark amplitudes of the
Bootstrap quark model. The main contribution describing the
interaction in the diquark channel (the states $J^P=0^+,1^+$) is
determined by gluon exchange in the t channel:
\begin{equation}
(\bar{q}_c\vec{\lambda}\gamma_{\mu}q)A(s,t)(\bar{q}_c\vec{\lambda}\gamma_{\mu}q)~~,
\label{eq:5}
\end{equation}
where $A(s,t)$ quark amplitude determines the strength of the
interaction of the quarks in the gluon channel. The mass $M_G$ of the
constituent gluon determines the region in which the gluon has a
short-range nature and has the consequence that the gluon field can
adjust to a change of the quark system and that non-potential effects
are successfully included. However, the quark amplitudes obtained in
the Bootstrap procedure depend not only on the square $t$ of the
momentum transfer, but also on the energy variable $s$. Therefore, a
literal transition to nonrelativistic potentials is not possible:
these amplitudes rather correspond to quasi-potentials. Then in
transforming from Bootstrap quark amplitudes to quark potential, the
energy $s=s_0$ is fixed, and already at fixed energy the dependence on
the momentum transfer is considered to be of a potential
nature. Fixing of the energy $s$ requires the introduction of a momentum
cut-off parameter $\Lambda_{\phi}$ in the Fourier transformation. This
parameter is chosen in such a way that the spectrum of low-lying
mesons and scalar diquarks calculated in the framework of the
Bootstrap procedure is reproduced. In our case we obtained
$\Lambda_{\phi}=5.55 ~m$, where $m$ is the mass of a constituent quark. 
The explicit form of the potentials of the quark interaction in a
color-antisymmetric state with quantum numbers $J^P=0^+,1^+$ for
non-strange quarks is given in Fig.\ref{fig:4}(a) and (b)~\cite{39}.

For the confinement potential we have chosen the usual linear
potential with a slope determined by the angle $\alpha$. We have
considered three cases, in which the confinement potential has been
added to the Bootstrap potential from the distances $r_0=0.6,1.0,1.3
~{\rm fm}$. The inclusion of such a potential shifts the spectrum of
low-lying mesons. In this case the nature of the change of the
spectrum is similar to the change of masses of the mesons and scalar
diquarks due to a change of the constituent quark masses. Therefore,
by inclusion of the confinement potential we have changed the masses
of the quarks so as to reproduce the spectrum of low-lying mesons. In
this case the Bootstrap potential was altered in the following way,
the scale of the abscissa was changed in proportion to $\sim 1/m$, and
the scale of the ordinate in proportion to\\$\sim m$.
In contrast to most quark potentials, the Bootstrap potential is
finite at $r=0$ because of the cut-off in the energy introduced in
the calculation of Bootstrap quark amplitudes. Moreover, the momentum
cut-off $\Lambda_{\phi}$ leads to small oscillations of the potential
at distances $\sim 1.5 ~{\rm fm}$, which are important for excited states
and do not have a significant effect on the spectrum of low-lying
baryons.\\
We have investigated S-wave baryons with the quantum numbers
$J^P=\frac{1}{2}^+,\frac{3}{2}^+$ consisting of u and d quarks. The
masses and wave functions of the baryons were obtained from the
solution of the Faddeev equations for the three-quark system with the
diquark potentials $V_{0^+}$ and $V_{1^+}$. Three-particle forces were
not included in this case. The results of the calculations are given
in Table~\ref{tab:8}.
\begin{table}[!h]
\begin{center}
\begin{tabular}{ccccccccccc}
$r_0$ & m & $\alpha$ & $m_N$ & $m_{\Lambda}$ & $\langle
R_p^2\rangle^{1/2}$ & $R_n^2$ & $\mu_p$ & $\mu_n$ & $(v/c)_N^2$ &
$(v/c)_{\Lambda}^2$ \\
fm & MeV & ${\rm GeV}^2$ & MeV & MeV & fm & $fm^2$ &  &  &  &  \\\hline
0.6 & 343 & 0.048 & 932 & 1236 & 0.37 & 0 & 2.73 & -1.82 & 1.3 & 1.1\\
1.0 & 336 & 0.138 & 944 & 1222 & 0.39 & 0 & 2.79 & -1.86 & 1.2 & 1.0\\
1.3 & 334 & 0.194 & 938 & 1190 & 0.41 & 0 & 2.81 & -1.87 & 1.2 & 0.96
\end{tabular}
\end{center}
\caption{\label{tab:8}Structure parameters of baryons for (I)-(III)
  versions of the potential, Fig.\protect{\ref{fig:4}}.}
\end{table}

The values of the masses of the nucleons and the $\Lambda$ isobar
agree well with the experimental values. In this case the mass of the
constituent u and d quarks is 336 MeV (version (II),Fig.\ref{fig:4}),
which gives good agreement with experiment for the magnetic moments
$\mu_p=2.79 (2.79)$, $\mu_n=-1.86 (-1.91)$. The charge radius of the
proton has been found to be a factor of two smaller than the
experimental value, $\langle R_p^2 \rangle^{1/2}=0.4~{\rm
  fm}~(0.71~{\rm fm})$.\\
The aim of this calculation is to investigate the possibility of
describing the lowest three-quark systems by means of a Bootstrap
potential, using the nonrelativistic Faddeev equations~\cite{40}. the
results of the calculations show that the spin-spin splitting of the
levels $J^P=\frac{1}{2}^+,~\frac{3}{2}^+$, which depends on the wave
function of the diquark at $r=0$, is reproduced rather well. Our
values of the mass of the constituent quarks are close to the standard
values. Therefore, the magnetic moment of the nucleons and the
$\Lambda$ isobar agree with the experimental values. The parameter
$v/c$ shows that we are dealing with a relativistic system of three
particles. Therefore, the next step must be to calculate the
characteristics of baryons in relativistic treatment.

\section{Relativistic three-quark equations and spectroscopy of
  low-lying baryons}
In the framework relativistic approach of three-hadron system one can
consider the pair interaction between the
particles~\cite{41}-\cite{43}. There are three isobar channels, each
of which consists of a two-particle isobar and the third particle. The
presence of the isobar representation together with the condition of
unitarity in the pair energies and analyticity leads to a system of
integral equations in a single variable. Their solution makes it
possible to describe the interaction of the produced particles in
three-hadron systems. \\
In our consideration we construct a relativistic generalization of the
three-particle Faddeev equations in the form of dispersion relation in
the pair energy of the two interacting particles. By the method of
extraction of the leading singularities of the amplitude we calculate
the mass spectrum of S-wave baryons, the multiplets
$J^P=\frac{1}{2}^+$ and $\frac{3}{2}^+$, and we obtained the electric
form factors of nucleons and hyperons at low and intermediate $Q^2$.
\begin{figure}[!h]
\begin{center}
\unitlength1cm   
\begin{minipage}[t]{90mm}
\epsfxsize=90mm\epsfbox{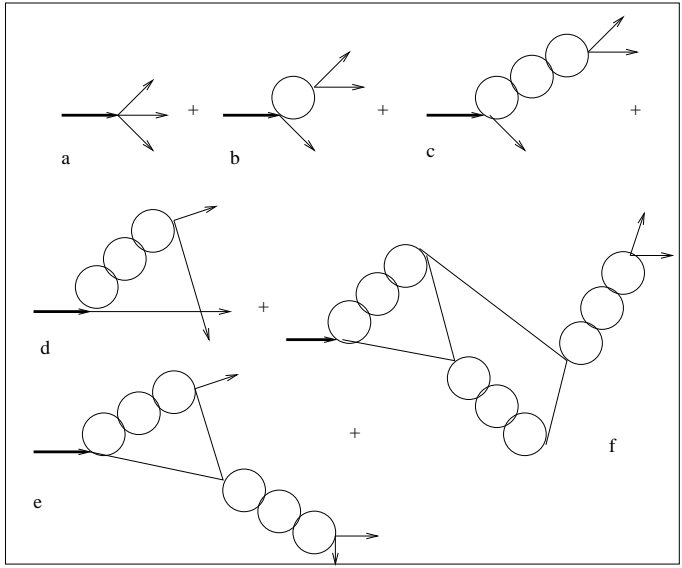}
\end{minipage}
\end{center}
\caption{\label{fig:5}Diagrams corresponding to: (a) production of
  three quarks, (b)-(f) successive pair interactions.}
\end{figure}

\begin{figure}[!h]
\begin{center}
\unitlength1cm   
\begin{minipage}[t]{90mm}
\epsfxsize=90mm\epsfbox{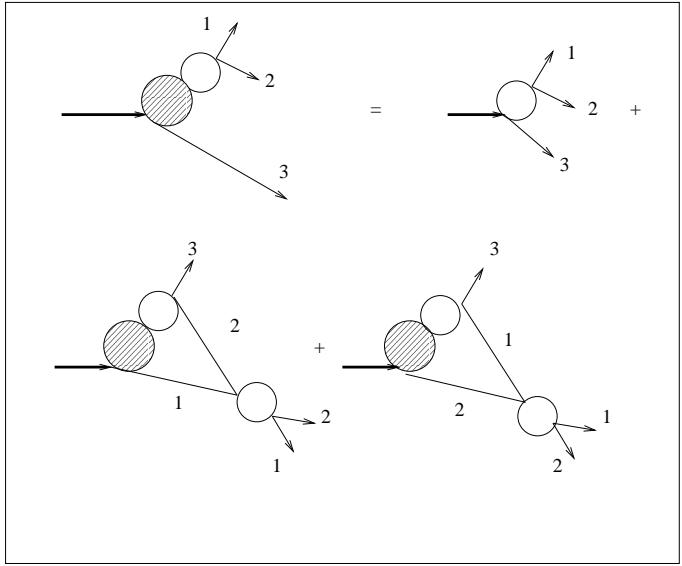}
\end{minipage}
\end{center}
\caption{\label{fig:6}Graphical representation of the equations for
  the amplitude interaction of three particles.} 
\end{figure}

The diagrams in Fig.\ref{fig:5} can be grouped according to which of
the three quark pairs undergoes the last interaction, i.e., the total
amplitude can be represented graphically as a sum of diagrams, as
shown in Fig.\ref{fig:6}.

Using the diagrams in Fig.\ref{fig:5} it is very easy to write down a
graphical equation for the amplitude interaction of three
particles. Then to write down a concrete equation we must specify the
amplitude of the pair interaction of the quarks. For this we shall use
the results of the Bootstrap quark model~\cite{26} and construct the
amplitude of the interaction of two quarks in the states $J^P=0^+$,
$1^+$. \\
The construction of the approximate solution of Fig.\ref{fig:6} is
based on the extraction of the leading singularities in the
neighborhood of $s_{ik}=4m^2$. The structure of
the singularities of 
amplitudes with a different number of rescatterings (Fig.\ref{fig:5})
is the following. The strongest singularities in $s_{ik}$ arise from
pair rescatterings of quarks $i$ and $k$: a square-root singularity
corresponding to a threshold and pole singularities corresponding to
bound states (on the first sheet in the case of real bound states, and
on the second sheet in the case of virtual bound states). The diagrams
of Fig.\ref{fig:1}~(b) and (c) have only these two-particle
singularities. In addition to two-particle singularities the diagrams
of Fig.\ref{fig:1}~(d) and (e) have their own specific triangle
singularities. The diagram of Fig.\ref{fig:1}~(f) describes a larger
number of three-particle singularities. In addition to singularities
of the triangle type it contains other weaker singularities. Such a
classification of singularities makes it possible to search for an
approximate solution of Eq.(Fig.\ref{fig:6}), taking into account a
definite number of leading singularities and neglecting the weaker
ones.\\
We use the approximation in which the singularity corresponding to a
single interaction of all three particles, the triangle singularity,
is taken into account. If we choose the approximation in which
two-particle and triangle singularities are taken into account, and if
all functions which depend on the pair energies will be determined at
the central point of the physical region of the Dalitz plot, the
problem of solving the system of integral equations reduces to one of
solving simple algebraic equations~\cite{44,45}.
In Table \ref{tab:9} we give the calculated masses of S-wave baryons.
\begin{table}[!h]
\begin{center}
\begin{tabular}{cccc}
  & $m(J^P=\frac{1}{2}^+)$ & &  $m(J^P=\frac{3}{2}^+)$ \\\hline
N        &   0.940 (0.940) &$\Delta$  &  1.232 (1.232) \\
$\Lambda$&   1.098 (1.116) &$\Sigma^*$&  1.377 (1.385) \\
$\Sigma$ &   1.193 (1.193) &$\Theta^*$&  1.524 (1.530) \\
$\Theta$ &   1.325 (1.315) &$\Omega^-$&  1.672 (1.672)
\end{tabular}
\end{center}
\caption{\label{tab:9}Baryon masses $m(J^P)$ (GeV) - Note: the
  parameters of the model are as follows: the cut-off
  $\lambda_1=12.2$, the cut-off strange diquarks with $J^P=0^+$
  $\lambda_0=9.7$, the vertex functions for $J^P=1^+$ and $J^P=0^+$
  diquarks, respectively ($g_1=0.540$, $g_0=0.702$).}
\end{table} 

\begin{figure}[!h]
\begin{center}
\unitlength1cm   
\begin{minipage}[t]{90mm}
\epsfxsize=90mm\epsfbox{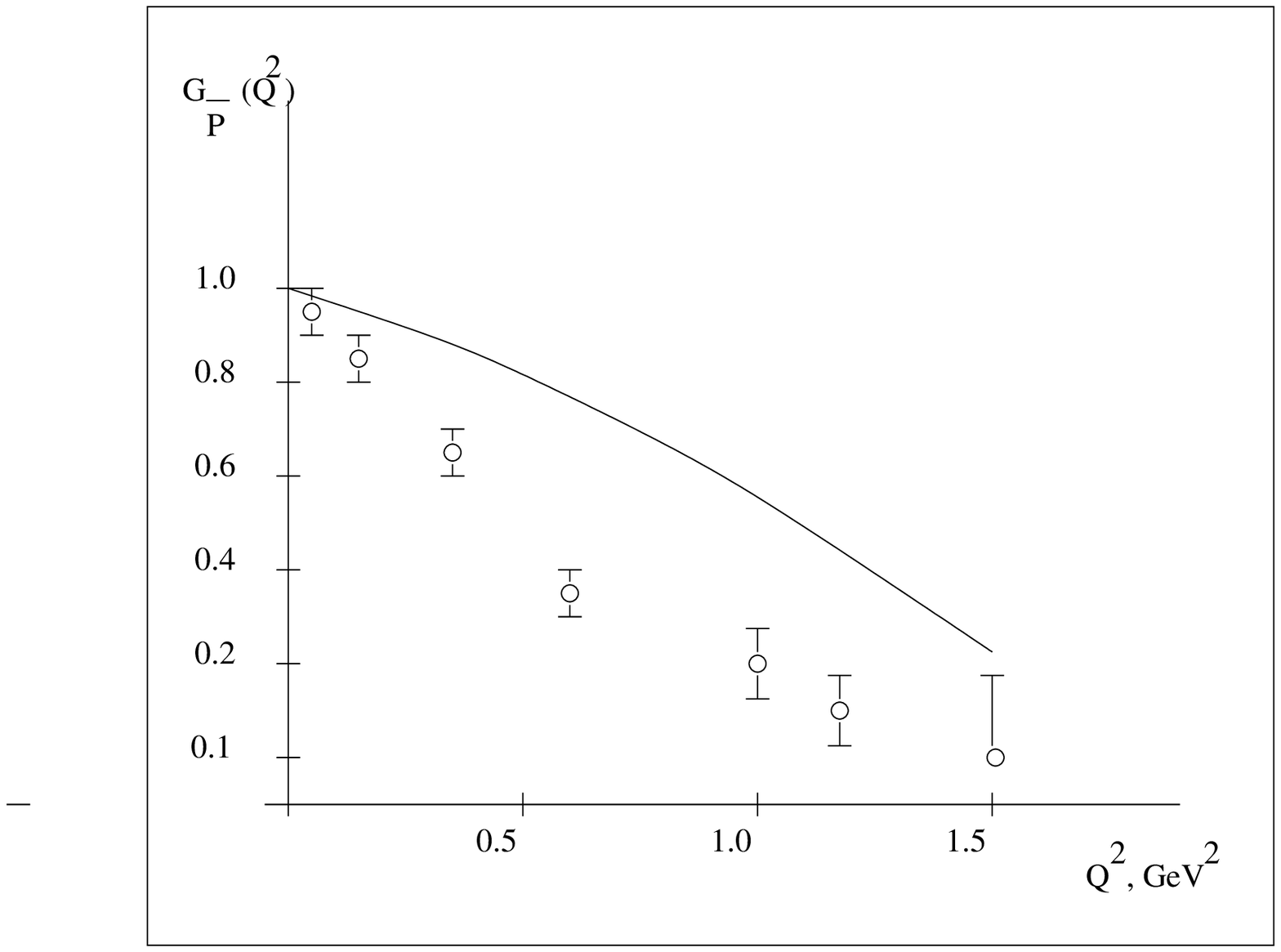}
\end{minipage}
\end{center}
\caption{\label{fig:7}Electric form factor of the proton at small and
  intermediate momentum transfer $Q^2 < 1.5 ~{\rm GeV}^2$ ($Q^2=-q^2$); the
  points are the experimental data~\protect{\cite{46}}.} 
\end{figure}

The calculated vertex functions $g_1$ and $g_2$ turned out to be close
to vertex function of diquarks in the Bootstrap procedure for S-wave
mesons. \\
In our approximate solution of the three-particle equations the
conditions of analyticity, unitarity and relativistic invariance of the
constructed three-particle amplitudes are satisfied. This
approximation is similar to the Bootstrap procedure for S-wave
mesons.\\
The dispersion technique makes it possible to determine the form
factors of composite particles (in our case, of baryons). On the one
hand, the technique of dispersion integration is relativistically
invariant and not related to the consideration of any specific
coordinate system. On the other hand, there are no problems with the
appearance of extra states, since in the dispersion relations the
contributions of intermediate states are under control. \\
The behavior of electric form factor of proton as a function of the
transverse momentum is shown in Fig.\ref{fig:7}. The form factor of
non-strange quark will be assumed to be the same for u and d quarks:
$f_q(q^2)=\exp(q^2/\Lambda_q)$ with $\Lambda_q=3 ~{\rm GeV}^2$ and 
$f_s(q^2)=\exp(q^2/\Lambda_s)$ with $\Lambda_s=5 ~{\rm GeV}^2$ for
s-quarks. We can not use new parameters as compared to Bootstrap procedure.

The calculated value of the charge radii of proton and hyperons was
found to be 
\begin{eqnarray}
\langle R_p^2 \rangle^{1/2} &=& 0.4 ~{\rm fm}~(0.8~{\rm fm})
\cite{46}\nonumber\\ 
\langle R_{\Sigma^+}^2 \rangle^{1/2} &=& 0.43 ~{\rm fm} \hspace{2em} 
\langle R_{\Sigma^-}^2 \rangle^{1/2} = 0.39 ~{\rm fm} \hspace{2em}
\langle R_{\Xi}^2 \rangle^{1/2} = 0.38 ~{\rm fm} 
\end{eqnarray}

The charge radius of the neutron was found to be practically equal to
zero. It is probable that only the inclusion of higher excitations can
lead to a result close to the experimental value ($R_N^2=-0.116
~{\rm fm}^2$~\cite{46}, $R_N^2=-0.012 ~{\rm fm}^2$~\cite{47}). One
points out that the new parameters for the calculations of electric
form factors of hyperons~\cite{48} did not use.

\section{Charmed baryons ($J^P=\frac{1}{2}^+,\frac{3}{2}^+$)}
In the framework of dispersion relation technique the relativistic
Faddeev equations for charmed baryons are found. The inclusion of
relativistic effects in composite system is important in considering
the quark structure of hadrons. We give the calculated masses of
S-wave charmed baryon (Table~\ref{tab:10})~\cite{49,50} and consider a
method of obtaining the electric form factors of charmed baryons with
quantum numbers $J^P=\frac{1}{2}^+$~\cite{51}.
\begin{table}[!h]
\begin{center}
\begin{tabular}{ccccc}
quark content &$J^P=\frac{1}{2}^+$ & M & $J^P=\frac{3}{2}^+$ & M \\\hline
udc         & $\Lambda^+_c$   & 2.284 (2.285)   & -       & -       \\      
uuc,~udc,~ddc &$\Sigma^{++,+,0}_c$& 2.458 (2.455)&$\Sigma^{*
  ++,+,0}_c$& 2.516 (2.519) \\
usc,~dsc      &$\Xi^{+,0(A)}_c$& 2.467 (2.467)&$\Xi^{* +,0}_c$&
2.725 (2.645) \\
usc,~dsc      &$\Xi^{+,0(S)}_c$& 2.565 (2.562)& -      & -        \\   
ssc           &$\Omega^0_c$ & 2.806 (2.704) & $\Omega^{* 0}_c$ & 3.108
(-) \\
ccu,~ccd      &$\Omega^{++,+}_{ccq}$& 3.527 (-)&$\Omega^{* ++,+}_{q}$&
3.597 (-)  \\ 
ccs           &$\Omega^{+}_{ccs}$& 3.598 (-)&$\Omega^{* +}_{s}$&
3.700 (-)  \\ 
ccc           & - & - &$\Omega^{++}_{c}$& 4.972 (-) 
\end{tabular}
\end{center}
\caption{\label{tab:10}Charmed baryon masses
  $J^P=\frac{1}{2}^+$,~$\frac{3}{2}^+$ (GeV) - in parenthesis the PDG
  data is presented~\protect{\cite{52}}; Note: parameters of model:
  cut-offs $\lambda_q=10.7$, $\lambda_c=6.5$ for q-~and c-quarks
  respectively, $g_c=0.857$ vertex function of charmed diquark. Masses
  quarks $m=0.495 ~{\rm GeV}$, $m_s=0.77 ~{\rm GeV}$, $m_c=1.655 ~{\rm
  GeV}$.}
\end{table}
The behavior of electromagnetic form factors of charmed baryons
($J^P=\frac{1}{2}^+$) in the region of low and intermediate momentum
transfers $Q^2 < 1.5 ~{\rm GeV}^2$ is determined. The calculated value
(without new parameters) of charge radii:
$\Lambda^+$,~$\Sigma^+_c$,~$\Xi^{+(\Lambda)}_c$,~$\Xi^{+(S)}$ are
equal $(0.12 - 0.20)$ fm, for the neutral baryons the charge radii are
equal to zero. 

\section{Conclusion} 
Calculations performed in the framework of Bootstrap quark model argue
in favor of the quasinuclear quark structure of hadrons. The resulting
quark interaction appeared to be effectively short-range. As was
stated above, this interaction is determined mainly by the exchange in
the gluon channel: the constituent-gluon mass appeared to be not
small. Moreover, calculations reveal an additional reason for the
short-rangeness of the interaction, connected with certain specific
features of the quark-quark interaction. The value of the constituent
gluon mass obtained in this model (700 MeV) seems to be rather
reasonable: just this mass value is required by hadron
phenomenology. The mass of the constituent gluon should be close to
that of vector particles. This is a consequence of
$1/N_c$-expansion~\cite{53,54}.   \\
Our calculation indicates an important role of interaction which is
induced by instantons. Such an interaction is necessary for deriving
both the pion mass and the $\eta - \eta'$ mass splitting. However,
the relative contribution of the instanton-induced interaction is less
than that with the gluon exchange, the ratio of forces about
 $1/4$. However, due to the rules of $1/N_c$-expansion this interaction
influence slightly the other channels while in the $\eta - \eta'$
channel it provides the correct values of masses and gives the angle
of $\eta_1 - \eta_8$ mixture close to that of the quark model. Just
the minor addition of ``instanton'' interaction produces the correct
value of the pion mass.\\
In the framework of the proposed approximate method of solving the
relativistic three-particle problem, we have obtained a satisfactory
spectrum of S-wave baryons. In this case it is obvious that the
interaction forces which determine the baryon spectrum are in fact the
same as for diquarks in the Bootstrap quark model for S-wave
mesons. On account of the rules of the $1/N_c$-expansion the diquark
forces are determined by one-gluon exchange (the instanton corrections
are small in this channel). One-gluon exchange corresponds to a
chromomagnetic interaction, which is responsible for the spin-spin
splitting.\\
The mass spectrum of charmed and beauty mesons and baryons is the
subject of many investigations.\\
In the framework of dispersion relation technique we calculated the
mass spectrum of charmed mesons and baryons. The calculations show
that the non-strange diquarks are in fact the same in the ordinary and
charmed baryons. The interaction of the heavy quarks is described by
quark amplitudes corresponding to exchange of light white and colored
mesons. The main role in the formation of the spectrum of heavy mesons
is played by the forces that correspond to exchange of a massive
gluon. Therefore we can point out, that the considered Bootstrap quark
model allows to describe the light and heavy hadron spectroscopy based
on three principles: unitarity, analyticity and relativistic symmetry
(crossing symmetry) in good agreement with experimental data.

\underline{Acknowledgment:}\\
The authors would like to thank Profs. V.A.~Franke, \\
Yu.V.~Novozhilov,
and H.C.~Pauli. S.M.~Gerasyuta thanks the Max-Planck Institut f\"ur
Kernphysik in Heidelberg for the hospitality where a part of this work
was completed.

\end{document}